\documentclass[twocolumn]{aastex701}

\usepackage{array}
\usepackage{enumitem}
\usepackage{amsmath}
\usepackage{subcaption}
\usepackage{tabularx}
\usepackage{graphicx}

\usepackage{amssymb}
\usepackage{mathtools} 
\usepackage{parskip}
\usepackage{comment}
\usepackage{multirow}
\usepackage{makecell} 
\usepackage{threeparttable}
\usepackage{hyperref}

\begin{document}

\title{The Subversive Role of Excessive External Shear in Concealing Lensing Anomalies}

\author[orcid=0000-0003-1276-1248,sname='Alfred']{Amruth Alfred}
\affiliation{Department of Physics, The University of Hong Kong, Pokfulam Road, Hong Kong}
\affiliation{The Hong Kong Institute for Astronomy and Astrophysics, The University of Hong Kong, Pokfulam Road, Hong Kong}
\email[show]{aalfred@hku.hk}  

\author[orcid=0009-0008-8500-7485,sname='Singh']{Shashpal Singh}
\affiliation{Department of Physics, The University of Hong Kong, Pokfulam Road, Hong Kong}
\email{shashpal130@gmail.com}

\author[orcid=0009-0004-2970-6805,sname='Lewis']{Rommulus Francis Lewis}
\affiliation{Department of Physics, The University of Hong Kong, Pokfulam Road, Hong Kong}
\affiliation{The Hong Kong Institute for Astronomy and Astrophysics, The University of Hong Kong, Pokfulam Road, Hong Kong}
\email{rommulus@connect.hku.hk}  

\author[orcid=0000-0003-3878-9118,sname='Chow']{Alex Chow}
\affiliation{Department of Physics, The University of Hong Kong, Pokfulam Road, Hong Kong}
\affiliation{The Hong Kong Institute for Astronomy and Astrophysics, The University of Hong Kong, Pokfulam Road, Hong Kong}
\email{alex4180@connect.hku.hk} 

\author[orcid=0000-0003-4220-2404,sname='Lim']{Jeremy Lim}
\affiliation{Department of Physics, The University of Hong Kong, Pokfulam Road, Hong Kong}
\affiliation{The Hong Kong Institute for Astronomy and Astrophysics, The University of Hong Kong, Pokfulam Road, Hong Kong}
\email{jjlim@hku.hk}

\author[orcid=0000-0003-3484-399X,sname='Oguri']{Masamune Oguri}
\affiliation{Center for Frontier Science, Chiba University, Chiba 263-8522, Japan}
\affiliation{Department of Physics, Graduate School of Science, Chiba University, Chiba 263-8522, Japan}
\email{masamune.oguri@chiba-u.jp}

\author[orcid=0000-0001-9065-3926,sname='Diego']{Jose M. Diego}
\affiliation{IFCA, Instituto de Física de Cantabria (UC-CSIC), Av. de Los Castros s/n, 39005 Santander, Spain}
\email{jdiego@ifca.unican.es}

\author[orcid=0000-0002-8785-8979,sname='Broadhurst']{Tom Broadhurst}
\affiliation{Donostia International Physics Center, DIPC, Basque Country, San Sebastián, 20018, Spain}
\affiliation{Department of Physics, University of Basque Country UPV/EHU, Bilbao, Spain}
\affiliation{Ikerbasque, Basque Foundation for Science, Bilbao, Spain}
    \email{tom.j.broadhurst@gmail.com}

%
\begin{abstract}

\noindent  To best reproduce observed multiply-lensed lensed images, lens models usually incorporate shear attributed to objects unrelated to the lensing galaxy (i.e., external shear): whether it be neighbouring galaxies not explicitly included in the lens model or other cosmic structures along the sightline.  When constrained solely by the positions of image counterparts, such lens models, even those utilising simple ellipsoidal mass distributions, can satisfactorily -- if not near perfectly -- reproduce the observed image positions, but often leave significant differences in flux ratios between the predicted and observed images.  For the narrow-line regions (NLRs) of quasars, which are too large to be affected by micro-lensing from stars in the lensing galaxy, the flux ratio anomalies thus left are commonly attributed to small-scale structures (sub-structures) in Dark Matter associated with the lensing galaxy. Here, we show that external shear can always resolve, among the quadruply-lensed quasar NLRs studied, position anomalies in lens models constrained solely by the observed image positions, and in addition reduce although not fully resolving flux ratio anomalies when constrained by both the observed image positions and flux ratios -- provided, usually, that the external shear incorporated have strengths that far exceed (as is the common practise) those typically inferred from weak lensing along general sightlines (i.e., cosmic shear).  Our work highlights the subversive role of excessive external shear in concealing lensing anomalies, undermining inferences on the characteristics of Dark Matter sub-structures -- and, correspondingly, the nature (mass and temperature) of the Dark Matter particle -- when not sensibly incorporated into lens models.

\end{abstract}



\section{INTRODUCTION} \label{introduction}
Historically, lens models that adopt simple (i.e., ellipsoidal) and smoothly-varying mass profiles for lensing galaxies often leave discrepancies between the predicted and observed flux ratios of multiply-lensed (specifically, quadruply-lensed) quasars \citep{Mao_1998, Metcalf_2002, Keeton_2003,wyn, Mao_2004, Kochanek_2004, Goldberg_2010,Xu_2015, Shajib_2018, nierenberg2019}. Micro-lensing by individual stars in the lensing galaxy can induce additional de-magnification or magnification given a suitable alignment and provided that the lensed source is sufficiently small, generating flux ratio anomalies without significantly altering the positions of multiply-lensed images.  In multiply-lensed systems involving the narrow-line regions (NLRs) of quasars, however, stellar micro-lensing is deemed ineffective at altering magnifications owing to the relatively large sizes of the emitting regions concerned -- making these systems particularly valuable for probing the nature of Dark Matter (DM).  In a landmark study of eight such quadruply-lensed systems by \citet{nierenberg2019}, doubling the then-known number of multiply-lensed quasar NLRs, simple and smooth-varying lens models were found to leave flux ratio anomalies in all these systems. \citet{nierenberg2019} attributed these anomalies to small-scale DM structures (i.e., DM sub-structure), for which the most commonly invoked sub-structures are DM sub-halos -- as predicted in abundance by cosmological simulations that adopt cool and ultra-massive particles (e.g., WIMPS) for DM.

\begin{figure*}[t]
    \centering
    \includegraphics[width=\textwidth]{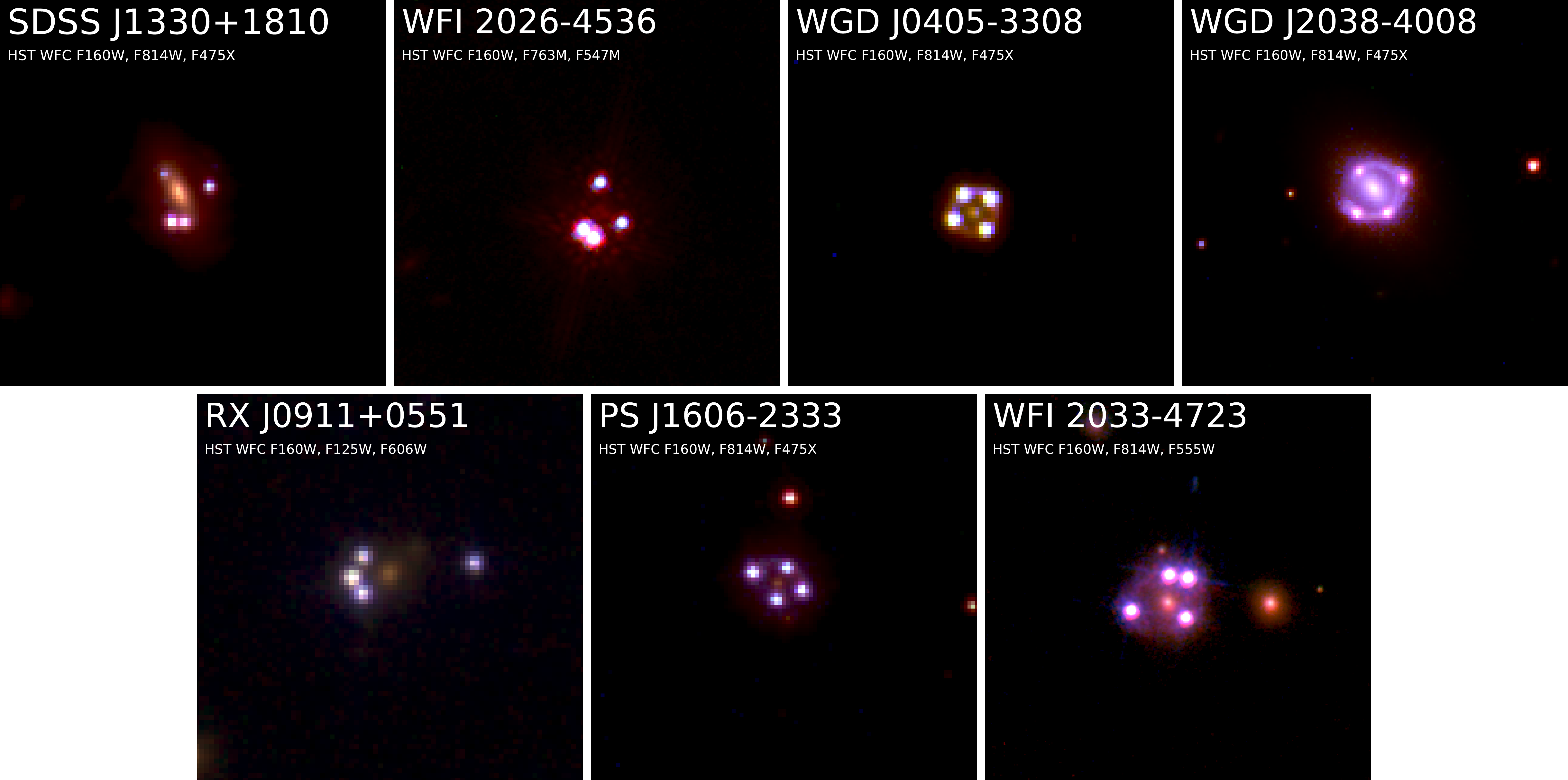}
    \caption{RGB images of the seven multiply-lensed quasars examined in this work, imaged with the Hubble Space Telescope.  Isolated systems (with no apparent neighbouring galaxies or not residing in a group or cluster environment) are displayed in the upper row, whereas non-isolated systems are displayed in the lower row.  PS\, J1606-2333 features a faint neighbouring galaxy (to the south of the lensing galaxy), whereas WFI\,2033-4723 and RX J0911+0551 reside, respectively, in group and cluster environments.}
    \label{fig:lenses}
\end{figure*}

The lens models constructed by \citet{nierenberg2019} are constrained solely by, and hence are designed to best reproduce, the positions of quadruply-lensed image counterparts.  This approach is commonly adopted by lens modellers under the presumption that DM sub-structures (in addition to, in cases where the lensed source is sufficiently small, stellar micro-lensing) can alter the magnifications of multiply-lensed images -- making the latter unsuitable for serving as constraints on simple and smoothly-varying lens models.   Lens models constructed in this manner, however, may disguise discrepancies in positions generated by DM sub-structures, and hence mislead inferences on their characteristics (e.g., masses and sizes of intervening DM sub-halos, if not a different type of sub-structure entirely such as the pervasive density modulations generated by cold and ultra-light DM particles owing to particle-wave interference).  In addition, such lens models \citep[e.g.,][]{Gilman_2019,gilman2024turbochargingconstraintsdarkmatter,Laroche_2022,Powell_2023,Amruth_2023} may overestimate the level of flux ratio anomalies as they are not designed specifically to reproduce the flux ratios of multiply-lensed images -- further misleading inferences on the characteristics of DM sub-structures.

In their work, \citet{nierenberg2019} assume that the lensing mass has a similar projected shape -- ellipticity and position angle -- as the light morphology of the lensing galaxy.  This approach is reasonable if the multiply-lensed images form at locations where the enclosed mass in stars dominates over that in DM, for which we lack a priori information.  As can be seen from Figure \ref{fig:lenses} \citep[and from Figure\,1 in][]{nierenberg2019}, the multiply-lensed NLR images often form far beyond the visibly bright regions of the lensing galaxy, where accordingly the enclosed DM mass may be an appreciable fraction if not dominate over the enclosed stellar mass.  In this situation, assuming that the lensing mass has a similar projected shape as the light morphology of the lensing galaxy then presupposes that the DM halo also shares a similar projected shape.

Finally, and perhaps most problematically, \citet{nierenberg2019} invoke relatively strong external shears in their lens models for all eight systems studied just to reproduce the quadruply-lensed NLR image positions.  This external shear may arise from (known or unknown) neighbouring galaxies not explicitly included in the lens model or an over-dense (i.e., group or cluster) environment, along with other cosmic structures along the sightline (the latter giving rise to cosmic shear).   In a landmark study that explored the role of external shear in simple galaxy lens models, \citet{1997Keeton} found that external shear was required to reproduce the positions of quadruply-lensed images among the systems studied -- but that the external shear strengths necessary are much higher (amounting to 10\%-15\% of the shear associated with the lensing galaxy itself) than those expected ($\sim$1\%-3\%) from cosmic shear.  Since then, more studies have addressed whether the external shear strengths incorporated into galaxy lens models are physical.  For example, from cosmological N-body simulations and semi-analytical models of galaxy formation, \citet{Holder_2003} found that lensing galaxies of  quadruply-lensed systems inhabit environments with stronger external shears -- other than cosmic shear -- than the values previously estimated: these relatively strong external shears stem from the tendency of early-type galaxies, which are the majority of lenses, to reside in overdense regions.  In a subsequent study populating cosmological N-body simulations with  massive elliptical galaxies, however, \citet{dalal} found that the external shear strengths associated with the environments of such lensing galaxies are far weaker than those invoked in galaxy lens models.  In a study of whether the external shears invoked for lensing galaxies of quadruply-lensed systems in group environments are consistent with those derived for these environments,  \citet{kenneth} find clear inconsistencies for half of the lenses  studied. Most recently, \citet{amy} highlighted that the external shears incorporated into lens models for galaxy lenses are almost always stronger than that inferred from independent weak lensing measurements towards the same systems.  We suggest that the lens models constructed by \citet{nierenberg2019} likely suffer the same problem for, at least, the relatively isolated systems in their sample: as a simple illustration of this issue, the external shear strengths invoked in their lens models for relatively isolated lensing galaxies overlap with those for lensing galaxies in groups or clusters.

Here, we make a more comprehensive exploration of simple and smoothly-varying lens models for seven of the systems (the remainder having poorly defined flux ratios owing to an overlapping pair of image counterparts) studied by \citet{nierenberg2019} and shown in Figure \ref{fig:lenses}, with a particular focus on the four relatively isolated lensing galaxies in this sample. By contrast with the work reported by \citet{nierenberg2019}, we begin by making two sets of lens models without incorporating external shear (as its strength is poorly constrained for any given system): those that are loosely constrained to follow the light morphology of the corresponding lensing galaxy versus those that are not constrained in this manner.  In this way, we can test whether relaxing morphological constraints on the lens models -- allowing for the possibility that the DM halo is not aligned with the light morphology and contributes significantly to the potential at the locations of the quadruply-lensed NLR images -- better reproduce the positions of the image counterparts, if not also their flux ratios.  As we will show, both these sets of lens models, even when constrained by the positions of the quadruply-lensed images alone, leave large position anomalies for all but one (involving a relatively isolated lensing galaxy) of the systems examined.

Next, we add external shear to both these sets of lens models (either designed to closely follow the light morphology of the lensing galaxy or freed from such considerations) to assess the external shear strengths required to best reproduce the positions, if not also flux ratios, of image counterparts.  In both sets of lens models, very strong external shears -- far exceeding those typically expected from cosmic shear -- are required for all but one of the isolated lensing galaxies just to reproduce the image positions alone.  Furthermore, as noted earlier for the lens models constructed by \citet{nierenberg2019}, the external shear strengths required in our lens models for the isolated systems overlap with those required in our lens models for the galaxy groups or clusters. Remarkably, when constrained by image positions alone, adding external shear often results in a near-perfect ability to reproduce the nominal positions of all four image counterparts, and can also (albeit not always) improve the ability to reproduce their flux ratios (albeit still leaving flux ratio anomalies for one or more  images).

Finally, we compare lens models constrained solely by the positions of the multiply-lensed NLR images with those constrained by both the positions and flux ratios of these images.  Once again, we make two sets of lens models: either those loosely restricted or freely unrestricted to follow the light morphology of their corresponding lensing galaxies.  In this situation, not unexpectedly, lens models trade off more poorly reproducing image positions with better reproducing image flux ratios.  Like before, adding external shear (of arbitrary strengths) always resolves position anomalies and can (although not always) reduce -- if not resolve -- flux ratio anomalies.  Our work therefore highlights the subversive role of external shear in concealing -- reducing if not resolving -- both position and flux ratio anomalies among simple and smoothly-varying lens models for quadruply-lensed quasars (more generally, unresolved multiply-lensed images).

\section{METHODOLOGY} \label{sec:methods}

Below we describe the ingredients for, the constraints placed on, and the metrics used to assess the performance of our suite of lens models for the seven (out of the eight) systems in \cite{nierenberg2019} that we studied. The  system excluded, HS 0810+2554, features two overlapping image counterparts, for which \cite{nierenberg2019} do not provide image fluxes. We focus in particular on the four isolated systems, while also constructing lens models for the three non-isolated systems to intercompare the shear strengths required for these two sets of systems. All the lens models were constructed using \textit{glafic} \citep{glafic1}, a parametric lens modelling software.

\subsection{ Lens Model Constraints}

In all previous work that looked at the systems we study here, only the positions of the quasar NLR image counterparts were used to constrain the lens models constructed -- all of which left behind flux ratio anomalies. Because image positions depend on the first derivative of the lensing potential, whereas image magnifications and therefore flux ratios depend on the second derivative of the lensing potential, it would not be surprising if lens models constrained by image position alone leave flux ratio anomalies. In our work, we construct a suite of lens models that are constrained solely by the positions of the quasar NLR image counterparts, as well as those constrained simultaneously by their positions and flux ratios. In the latter situation, \textit{glafic} weights all constraints equally in its minimization algorithm.  We conducted Markov Chain Monte Carlo simulations to ensure that our best-fit lens models correspond to a global minimum in best-fit solutions.  

\subsection{ Lens Model Ingredients}

    We try three different mass profiles for representing the lensing potential of the lensing galaxy: (i) the conventionally used singular isothermal ellipsoid (SIE); (ii) a general power law (POW) with a density slope that is free to vary between -1.3 to -2.5; and (iii) a Navarro-Frenk-White (NFW) profile, which has a shallower density slope of -1 at the inner regions changing to a steeper slope of -3 at the outer regions \citep{Zhao_1996}. An SIE is a special case of a power law with a density slope of -2, widely adopted as a reasonable approximation to the dark matter and baryonic (primarily stellar) mass components combined.  This approach can  be traced back to work by \citet{Treu_2004}, who employed both strong lensing and measurements of stellar velocity dispersion to infer nearly (somewhat flatter than) isothermal profiles for early-type galaxies (within their inner few effective radii), albeit with considerable scatter among the different lensing galaxies studied. For the DM halo alone, cosmological hydrodynamical simulations predict an NFW profile, which can be a good representation of the lensing potential in cases where lensed images form far away from the stars.

As mentioned above, the systems we examined include three non-isolated lensing galaxies, which in one case features a neighbouring galaxy (PSJ1606). For the latter, we used an SIE to parameterise the contribution of the neighbouring galaxy to the total lensing potential.   The other two non-isolated galaxies lie, respectively, in a group and cluster environment. For all seven systems studied, we constructed lens models with and without external shear, which is centered on the lensing galaxy and parameterised by an amplitude as well as a position angle. 

To ensure that our model lens centre does not deviate too much from the inferred light centre of the (main) lensing galaxy, we adopt a Gaussian prior with a standard deviation of 0\farcs01 centered  on the nominal light center of this galaxy.  We adopt the same prior for the center of the SIE component representing the neighbouring galaxy in PSJ1606. The nominal light centers of the respective galaxies were determined by \citet{nierenberg2019}; the standard deviation adopted for the prior corresponds to the measurement uncertainty quoted by \citet{nierenberg2019} for the light centers of these galaxies.

For the lens models that are constrained by the light morphology of the (main) lensing galaxy, we adopt uniform priors for their ellipticities ranging from 0 to 0.2 for the circular lensing galaxies, and from 0.2 to 0.6 for the elliptical lensing galaxies. We deem this range to be a reasonable parameter space to search, as the lensing galaxies do not appear to be extremely elliptical in the HST images (\cite{nierenberg2019} find, for their best-fit lens models, ellipticities spanning the range 0--0.2). For position angles, we set no priors for the circular lensing galaxies, and a uniform prior within $\pm$15 degrees of the inferred position angle of the light image for the elliptical lensing galaxies. The priors adopted allow for a mild difference in shape and a mild misalignment in the position angle between the lens model and the stellar light. For the set of lens models that are not constrained by the observed light morphology of the lensing galaxy, we do not set any priors on the model parameters except for the lens centre as mentioned above.

\subsection{Performance Metrics}

The performance of each lens model is assessed by calculating: (i) the differences between the predicted and observed positions of the individual image counterparts; and (ii) the ratios between the predicted and observed fluxes for each pair of image counterparts (referenced to the brightest image counterpart). Following the common approach, we also calculated the root-mean-square (rms) dispersion between the predicted and observed positions as well as flux ratios for all four image counterparts to obtain simpler, but less detailed, performance metrics. We emphasize, however, that rms dispersions can mask relatively large lensing anomalies for an individual image counterpart if the remaining image counterparts are well reproduced --- highlighting the importance of examining performance metrics for individual image counterparts, as we focus on in our work.

\section{RESULTS} \label{sec: result}
We focus our presentation of the results on the four relatively isolated systems contained in \citet{nierenberg2019}: SDSS J1330, WFI 2026, WGD J0405 and WGD J2038 hereafter, with their full names as shown in Figure \ref{fig:lenses}. No neighbouring galaxies have been identified by \citet{nierenberg2019} in any of these systems, nor are any visibly apparent based on our visual inspection of their images. Although we do not present detailed results for the three remaining systems (full names as shown also in Fig.\ref{fig:lenses}), which in one case features a neighbouring galaxy (PSJ1606) and in the other two cases are associated, respectively, with a group (WFI2033) and cluster (RXJ0911), the external shear strengths required to resolve position anomalies in these systems serve as instructive comparisons with the external shear strengths required to resolve position anomalies in the isolated systems. Lens models constrained solely by the positions of image counterparts are presented first in Sect \ref{subsec:1} and \ref{subsec:2}, and then those constrained by both the positions and flux ratios of image counterparts presented in Sect \ref{subsec:4}. For simplicity in presentation, we focus here on the lens models constructed using an elliptical power-law (POW) profile, which provide the closest agreement to the observational constraints imposed on the individual lens models: the parameters of all the lens models constructed can be found in a repository (\url{https://github.com/romms921/SubversiveExternalShear}), with a select few listed in Table \ref{table1}. The lessons we draw from these lens models also apply to those constructed using either an SIE or NFW profile.

\subsection{Guided by Light Morphology} \label{subsec:1}

\begin{figure*}[t]
    \centering
    
    \begin{subfigure}[b]{0.49\textwidth}
        \centering
        \includegraphics[width=\textwidth]{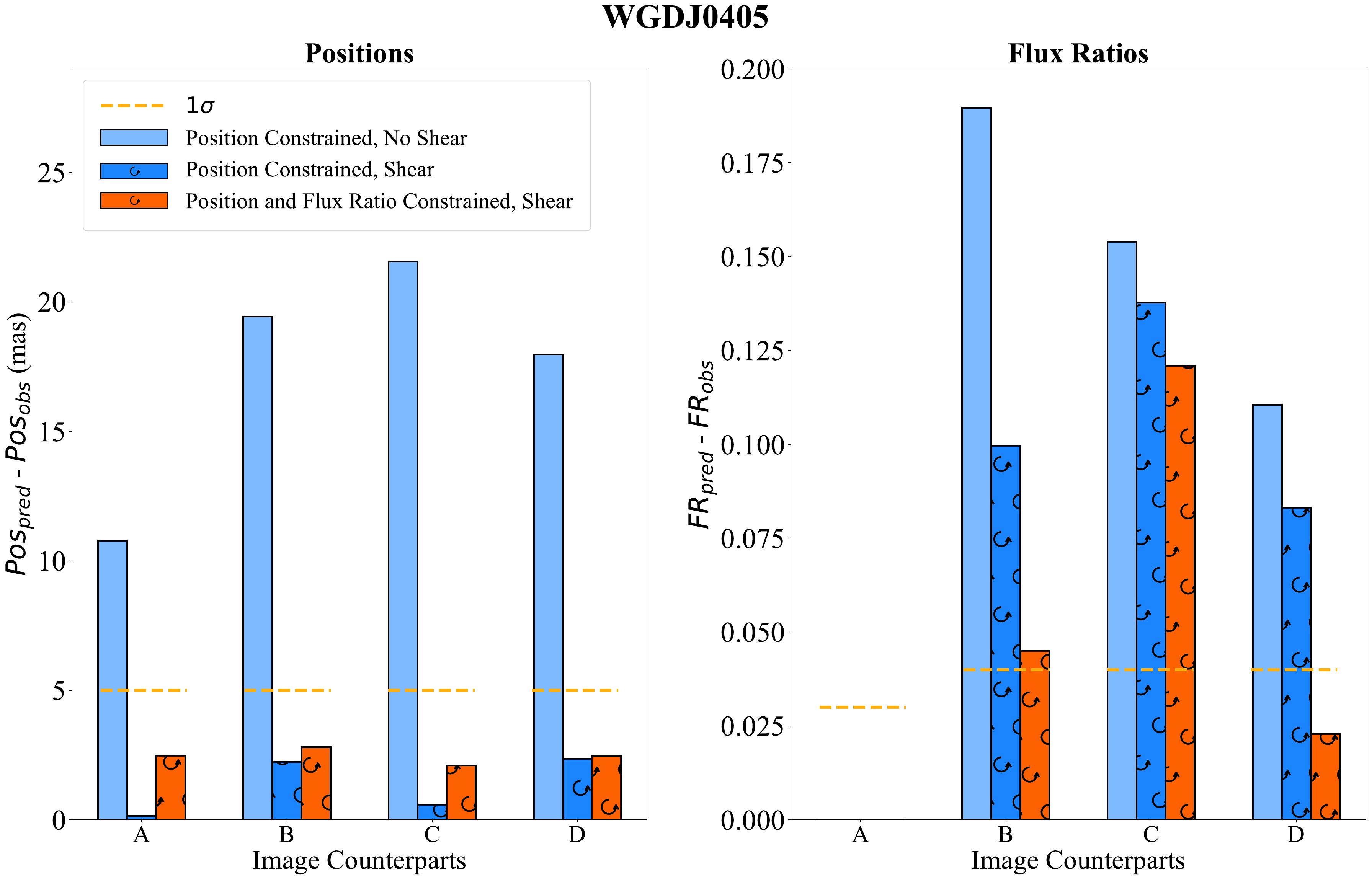}
        \label{fig:sub1}
    \end{subfigure}
    \hfill 
    \begin{subfigure}[b]{0.49\textwidth}
        \centering
        \includegraphics[width=\textwidth]{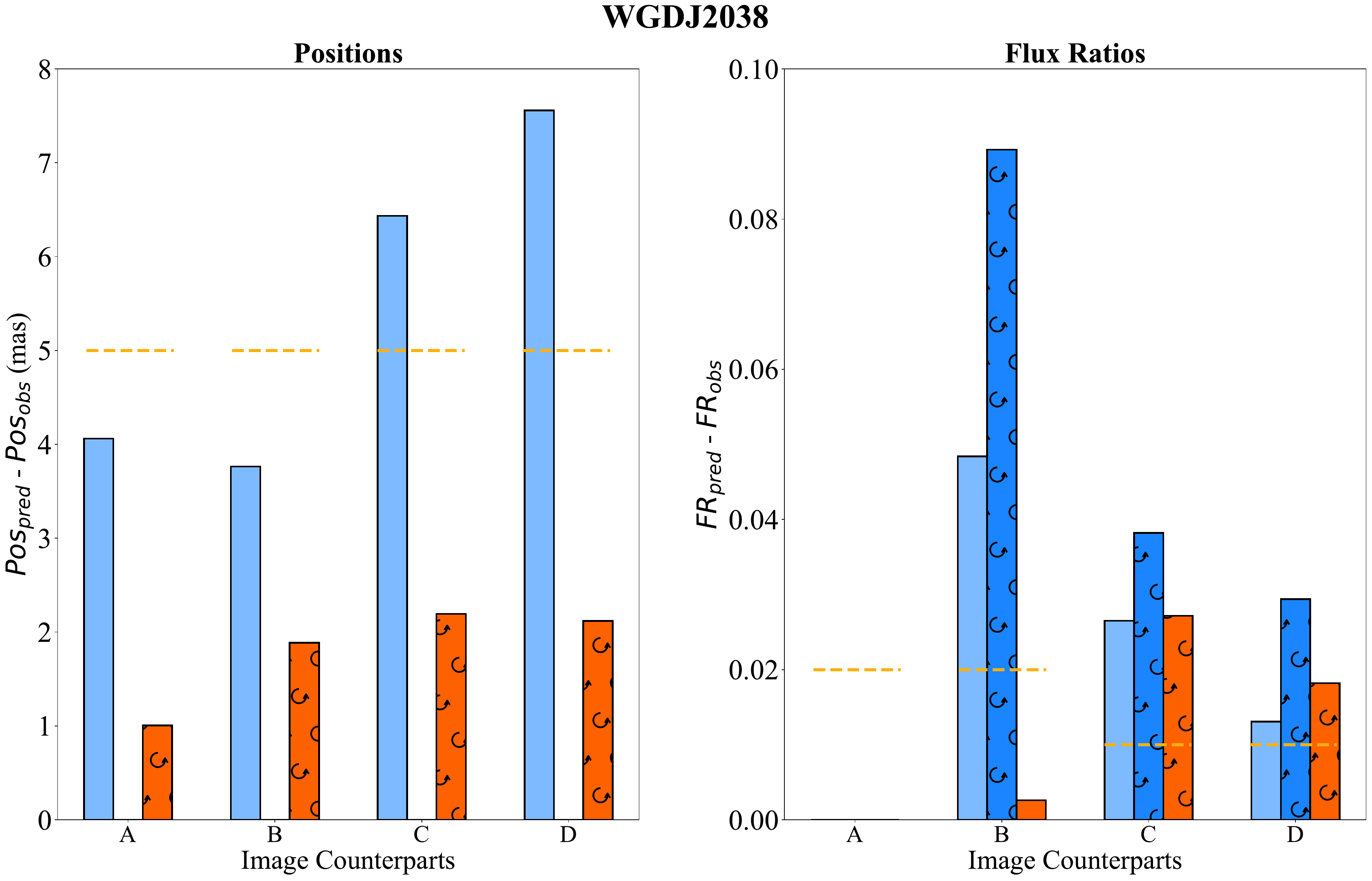}
        \label{fig:sub2}
    \end{subfigure}
       \caption{Illustrative examples of differences left by our best-fit power-law elliptical lens models between predicted and observed positions and flux ratios for two isolated systems.  Flux ratios are computed with respect to the brightest image counterpart (image A).  Lens models are constrained either by positions of four image counterparts alone (blue bars) or both their positions and flux ratios (red bars).  Lens models incorporating external shears are indicated by curly arrows.  Dash horizontal lines are measurement uncertainties on the observed image counterparts.  Lens models not incorporating external shear usually, albeit not always, leave large positions anomalies, as seen for WGDJ0405.  Adding external shear (having unconstrained strengths) brings often near-perfect (e.g., for WGDJ2038) agreement between predicted and observed positions of all four image counterparts; and often, albeit not always, also reduce if not resolve flux ratio anomalies for one or more image counterparts, especially for lens models constrained by both image positions and flux ratios.}
    
    \label{fig:main}
\end{figure*}

We start by constructing lens models that are guided by the light morphology of the lensing galaxies, as in the work of \citet{nierenberg2019}.  We find that the best-fit lens models thus constructed for three of the isolated systems leave significant ($\geq 3 \sigma$) position anomalies when no external shear is invoked.  By contrast, even without invoking external shear in the lens model for the remaining isolated system (WGD2038), the predicted and observed image positions are discrepant at just above $1 \sigma$ for two image counterparts and slightly below $1 \sigma$ for the other two image counterparts. Nonetheless, the lens model for this system leaves appreciable ($\sim 2\sigma$) flux ratio anomalies for two image counterparts. As illustrative examples, Figure \ref{fig:main} shows the level of agreement between the predicted and observed positions and flux ratios of individual image counterparts based on the best-fit lens models for two systems (WGDJ0405 and WGDJ2038).

When we allow for external shear with no restrictions on its strength, \textit{all} the lens models now predict  positions that are less than $1 \sigma$ discrepant if not in near-perfect agreement with the nominal positions for \textit{all four} image counterparts, as shown by the examples in \autoref{fig:main}. In the single case for which the predicted and observed image positions are already near satisfactory agreement ($\sim$1$\sigma$) even without invoking external shear (WGDJ2038), only a relatively weak external shear (commensurate with what might be expected from cosmic shear) is required to bring the positions to near-perfect agreement. For the other three isolated systems, however, the required shear strengths are very large (comparable to those required in the lens models constructed by \citet{nierenberg2019} for the corresponding systems) and overlap in range with those required for satisfactory image position agreements in the three non-isolated systems. Somewhat unexpectedly  for these three isolated systems, despite not using image flux ratios as constraints, their flux ratio anomalies are reduced with discrepancies now ranging from $\sim$$1\sigma$--$3 \sigma$, as shown by the example in  \autoref{fig:main}. On the other hand, for WGDJ2038, the predicted image flux ratios become even more discrepant (now exceeding $3 \sigma$ for one image counterpart) from those observed, as shown in the same figure.  Adding external shear can therefore often, albeit not always, reduce if not resolve flux ratio anomalies for one or more image counterparts even in lens models constrained by image positions alone.

\begin{table}[hbt!]
\begin{threeparttable}
\caption{Power-law Elliptical Lens Model Parameters}
\label{table1}

\footnotesize

\begin{tabular*}{\columnwidth}{@{\extracolsep{\fill}} lllllll @{}}
\hline
\hline
System & Light\tnote{a} & $e$ \tnote{b} & $PA(^\circ$)\tnote{c}& Index\tnote{d} & $\gamma_{\rm str}$\tnote{e}& $\gamma_{\rm PA}$\tnote{f}\\ 
 & Guided & & & & & \\
\hline
\multirow{5}{*}{SDSSJ1330} & \multirow{2}{*}{No} & 0.118 & 146.4  & 1.60  & -  & - \\
 &  & 0.570 & -21.6 & 2.34  & 0.076 & 101.9 \\
 &  &  &  &  &  &  \\
 & \multirow{2}{*}{Yes} & 0.222 & -33.1 & 1.90  & - & - \\
 &  & 0.517 & -21.9 & 2.19  & 0.078 & 93.9 \\
\hline
\multirow{5}{*}{WFI2026} & \multirow{2}{*}{No} & 0.265 & 90.1 & 1.70  & - & -\\
 &  & 0.489 & 77.9 & 2.19  & 0.079 & 115.0 \\
 &  &  &  &  &  &  \\
 & \multirow{2}{*}{Yes} & 0.200 & 89.4 & 1.90  & - & - \\
 &  & 0.200 & 72.5 & 2.11  & 0.107 & 100.1 \\
\hline
\multirow{5}{*}{WGDJ0405} & \multirow{2}{*}{No} & 0.064 & -30.8 & 1.70  & - & - \\
 &  & 0.243 & -43.5 & 2.16  & 0.035 & 16.2 \\
 &  &  &  &  &  &  \\
 & \multirow{2}{*}{Yes} & 0.196 & -30.8 & 2.14  & - & - \\
 &  & 0.200 & -46.5 & 2.04  & 0.038 & 17.4 \\
\hline
\multirow{5}{*}{WGDJ2038} & \multirow{2}{*}{No} & 0.341 & -35.4 & 1.90  & - & - \\
 &  & 0.319 & -37.0 & 1.90  & 0.011 & -13.7 \\
 &  &  &  &  &  &  \\
 & \multirow{2}{*}{Yes} & 0.365 & -35.3 & 1.93  & - & - \\
 &  & 0.319 & -37.0 & 1.90  & 0.011 & -13.8 \\
\hline
WFI2033\tnote{g} & \multicolumn{3}{c}{Range of $\gamma_{str}$} & \multicolumn{3}{c}{0.036 - 0.168}\\ 
RXJ0911\tnote{h} & \multicolumn{3}{c}{Range of $\gamma_{str}$} & \multicolumn{3}{c}{0.079 - 0.314}\\ 
\hline
\hline
\end{tabular*}

\begin{tablenotes}
\footnotesize 
\item[a] Guided or freely unconstrained by  light morphology of lensing galaxy
\item[b] Ellipticity of lens model
\item[c] Position angle of major axis of lens model
\item[d] Power-law exponent of lens model
\item[e] External shear strength (dash indicates no external shear incorporated into lens model)
\item[f] Position angle of external shear (centred on lensing galaxy)
\item[g] Group environment
\item[h] Cluster environment
\end{tablenotes}
\end{threeparttable}
\end{table}

\subsection{Not Guided by Light Morphology}\label{subsec:2}

We now ask whether the required external shear strengths would be more realistic if we allow greater freedom by not requiring the lens morphology to closely follow the light morphology. Permitting this greater flexibility allows for the lensing potential to be affected or dominated by a DM component that has a different shape from and/or is misaligned with the stellar component. We therefore constructed lens models with no priors on their shapes, although still requiring their centres to closely coincide with the light centers of the lensing galaxies as before.

When no external shear is invoked, all these lens models still leave position anomalies at levels comparable to those left by the corresponding set of lens models that are required to quite closely follow the light morphology of the lensing galaxy. The parameters of these lens models, listed in Table \ref{table1}, differ little from those guided by the light morphology of the lensing galaxy and also not incorporating external shear.

    On the other hand, when external shear is incorporated into the lens models, the predicted and observed image positions are -- like before -- brought into adequate ($\sim$$1\sigma$) if not near-perfect agreement for all four isolated systems, as illustrated by the two examples in \autoref{fig:main}. 
    As can be seen in Table \ref{table1}, the lens model parameters are mostly (although not always) very similar to the corresponding lens models where the lens morphology is required to quite closely follow the light morphology of the lensing galaxy - and therefore also the external shear strengths required in both these sets of lens models. Finally, like the lens models guided by the light morphology of the lensing galaxy, all these lens models leave significant ($\geq3\sigma$) flux ratio anomalies for at least one image counterpart.

\subsection{Adding Flux Ratio Constraints} \label{subsec:4}

Including flux ratios as an additional constraint results in a trade-off between better reproducing the flux ratios at the cost of more poorly reproducing (albeit not necessarily leaving unsatisfactory agreement in) the positions of image counterparts. Not surprisingly therefore, without invoking external shear, such lens models continue to leave position anomalies in all four isolated systems. 

Invoking external shear in lens models constrained by both positions and flux ratios leaves no position anomalies; indeed, like the lens models constrained by just positions alone, these lens models provide satisfactory (leaving at most $\sim$$1\sigma$ differences) if not near-perfect positional agreements for all image counterparts in all four isolated systems. Interestingly, all of these lens models -- irrespective of whether they are required to quite closely follow the light morphology of the lensing galaxy or not -- leave no flux ratio disagreements larger than $\sim$3$\sigma$.  Indeed, for most of the image counterparts, the disagreements are at the level of $\sim$1$\sigma$, and in only one system (WGDJ0405) is a single image counterpart discrepant in flux ratio at the 3$\sigma$ level as can be seen in \autoref{fig:main}; nonetheless, all such lens models leave discrepancies in flux ratios of at least $2\sigma$ for one or more image counterparts. The external shear strengths required are comparable to those of the lens models reported earlier satisfying just the positional constraints, and span the same range as the shear strengths necessary for the non-isolated systems satisfying the same constraints.

In brief, not only can an unrealistically strong external shear artificially eliminate position anomalies (indeed, bring often near-perfect agreements between the predicted and observed positions for all image counterparts), but  also artificially alleviate although not fully resolve flux ratio anomalies (leaving differences in flux ratios of $\gtrsim 2\sigma$ for at least one image counterpart)  in lens models constrained by both the positions and flux ratios of  compact quadruply-lensed images. These findings highlight the subversive role of excessive external shear in concealing lensing anomalies among such systems.

\section{SUMMARY} \label{sec:floats}

We demonstrated how smoothly-varying elliptical lens models that invoke overly strong external shears, a  common practise,  always vastly reduce if not completely eliminate position anomalies and can also reduce if not resolve flux ratio anomalies among compact multiply-lensed images.  Specifically, for quadruply-lensed systems, we show that: (i) no matter what density profiles (power-law or NFW profiles) we try, and even with no priors on the shapes and orientations of these lenses, we often need to invoke very strong external shears just to satisfactorily reproduce the positions of image counterparts; (ii) lens models constrained by image positions alone, when incorporating external shears of arbitrary strengths, always reproduce -- to near perfection -- the positions of all image counterparts; (iii) not only can incorporating external shear resolve position anomalies, but also reduce flux ratio anomalies even when flux ratios are not used as constraints on lens models; (iv) when constrained by both positions and flux ratios, incorporating external shear having arbitrary strengths can further reduce although not fully resolve flux ratio anomalies (leaving differences in flux ratios of $\gtrsim 2\sigma$ for at least one image counterpart) while still satisfactorily reproducing image positions; and (v) the external shear strengths necessary in lens models for relatively isolated lensing galaxies overlap in range with those invoked for lensing galaxies in groups and clusters.  Our work therefore highlights the subversive role of excessive external shear in concealing -- either greatly reducing or resolving -- both position and flux ratio anomalies among compact quadruply-lensed systems. 

The narrow-line regions (NLRs) of quasars, the quadruply-lensed features studied in our work, are too large to be affected by stellar micro-lensing, which can significantly alter the magnifications but not positions of image counterparts.  For these systems, smoothly-varying elliptical lens models that incorporate external shears having arbitrary strengths, and constrained by the positions of image counterparts alone, commonly leave flux ratio but not position anomalies.  Such flux ratio anomalies are commonly attributed to small-scale structures (sub-structures) in Dark Matter associated with the lensing galaxy.   Incorporating excessive external shear, however, misleads inferences on the type and level of lensing anomalies left by smoothly-varying elliptical lens models, and in consequence also deductions made about the characteristics of Dark Matter sub-structures -- thus undermining conclusions reached about the nature (mass and temperature) of the Dark Matter particle.

Expanding on multiply-lensed quasar NLRs, \citet{nierenberg2023jwstlensedquasardark} have reported a JWST survey of the warm-dust regions of 31 multiply-lensed quasars.  Like quasar NLRs, the warm-dust regions of quasars also are too large to be affected by stellar micro-lensing.  The trends found as summarised above can therefore be tested on a much larger sample of quadruply-lensed quasars.  Using power-law elliptical lens models with both external shear and angular multipoles to reproduce the multiply-lensed warm-dust regions of these quasars, \citet{keeley2024jwstlensedquasardark,keeley2025jwstlensedquasardark} and \citet{gilman2025b} place constraints on the particle mass for warm dark matter. We caution that all such work should pay attention to whether excessive external shear is being invoked in lens modelling, which if the case compromises the conclusions reached.

In a future work \citep{rommulus_shear}, we show that allowing external shears of arbitrary strengths results in a degeneracy between model parameters for power-law elliptical lenses and external shear strengths, making all such lens models unreliable.

\begin{acknowledgments}
A.A., S.S., R.F.L., A.C. and J.L. acknowledge support from the Research Grants Council of Hong Kong under the General Research Fund grant No. 17304425. We also acknowledge support from the Seed Fund for Collaborative Research from the University of Hong Kong.

Some of the computations in this paper were performed using research computing facilities offered by Information Technology Services at the University of Hong Kong.

The observations of the HST RGB images of the seven lensed quasar systems are associated with the programs \#15320, \#12874, \#8705 and \#9744.
\end{acknowledgments}

\bibliography{Bibliography}{}
\bibliographystyle{aasjournalv7}



\end{document}